\documentclass[traditabstract]{aa}
\usepackage{txfonts}

\usepackage{natbib}
\bibpunct{(}{)}{;}{a}{}{,} 

\usepackage{graphicx}
\usepackage{amssymb}
\usepackage[hyphens]{url}

\begin{document}

\title{\emph{XMM-Newton} RGS observations of the Cat's Eye Nebula}
\author{
M.A.\,Guerrero$^{1}$, J.A.\,Toal\'{a}$^{1}$, Y.-H.\,Chu$^{2,3}$, \and 
R.A.\,Gruendl$^{2}$}

%NOTA: \and para el ultimo autor de la lista

\institute{
$^{1}$Instituto de Astrof\'\i sica de Andaluc\'\i a, IAA-CSIC, Glorieta 
de la Astronom\'\i a s/n, 18008 Granada, Spain; \underline{mar@iaa.es}\\ 
$^{2}$Department of Astronomy, University of Illinois, 1002 West Green 
Street, Urbana, IL 61801, USA\\
$^{3}$Now at Institute of Astronomy and Astrophysics, Academia Sinica, 
Taipei, Taiwan, R.O.C.
}

\abstract{
We present an analysis of \emph{XMM-Newton} Reflection
  Grating Spectrometer (RGS) observations of the planetary nebula (PN)
  NGC\,6543, rendering it the second PN with high resolution X-ray
  spectroscopic observations besides BD\,$+$30$^{\circ}$3639.  The
  observations consist of 26 pointings, of which 14 included RGS
  observations for a total integration time of 435 ks.  Many of these
  observations, however, were severely affected by high-background
  levels, and the net useful exposure time is drastically reduced to 25
  ks.  Only the O~{\sc vii} triplet at 22 \AA\ is unambiguously
  detected in the RGS spectrum of NGC\,6543.  We find this spectrum
  consistent with an optically thin plasma at 0.147 keV (1.7 MK) 
and nebular
  abundances.  Unlike the case of BD\,$+$30$^{\circ}$3639, the X-ray
  emission from NGC\,6543 does not reveal overabundances of C and Ne.
The results suggest the N/O ratio of the hot plasma is consistent 
with that of the stellar wind, i.e., lower than the nebular N/O 
ratio, but this result is not conclusive.  
}


\titlerunning{\emph{XMM-Newton} RGS view of the Cat's Eye Nebula}
\authorrunning{Guerrero et al.}  

\maketitle

%%%%%%%%%%%%

\section{Introduction}

Since the \emph{EXOSAT} discovery of soft X-ray emission from
NGC\,1360 \citep{deKorte1985}, planetary nebulae (PNe) have been
routinely targeted by all subsequent X-ray missions.  Those first
detections, attributed to the photospheric emission of hot central
stars (CSPNe), were soon followed by reports of harder X-ray emission
originating in hot plasmas \citep[e.g.,][]{Leahy1994,Arnaud1996}.  The
X-ray observations of PNe up to the \emph{ROSAT} era
\citep{Guerrero2000} set the scene for the major leap in the study
of their hot gas content made with the advent of \emph{Chandra} and
\emph{XMM-Newton} \citep[see][and references
  therein]{Kastner2012,Ruiz2013}.
% It is now recognized two sources of X-ray emission in PNe: 
% diffuse emission from hot plasma filling the innermost cavity 
% and point-like emission from the CSPN.  

The production of hot gas is a fundamental prediction of the 
Interacting Stellar Winds (ISW) models of PN formation and 
evolution \citep{Kwok1978,Balick1987}.  
According to these models, the fast stellar wind developed during the 
post-asymptotic giant branch (AGB) sweeps up the previous slow and dense 
AGB wind.  
As a reverse shock propagates into the fast stellar wind, the central 
cavity is expected to be filled with highly pressurized shocked fast 
wind at temperatures of 10$^7$--10$^8$ K.  
This hot gas is too tenuous to produce appreciable X-ray emission, but 
the mixing of cool nebular material into the hot interior at their 
interface lowers the temperature and raises the density of the hot 
interior gas, producing optimal conditions for soft X-ray emission 
\citep{Steffen2008,ToalaArthur2014}.  
The detection of diffuse X-ray emission, first hinted by \emph{ROSAT} 
\citep{Kreysing1992}, then finally resolved by \emph{Chandra} 
\citep{Kastner2000,Chu2001}, marked a milestone in our understanding 
of the formation and evolution of PNe.

ChanPlaNS, the \emph{Chandra} Planetary Nebula Survey, is producing 
an unbiased view of the X-ray emission from PNe within 1.5 kpc.  
This survey is revealing notable correlations between the diffuse 
X-ray emission of PNe and their nebular and stellar properties 
\citep{Freeman2014}, but these results are based on low-resolution 
CCD imaging spectroscopic observations, which cannot provide a 
comprehensive description of the physical conditions and chemical 
abundances of the hot gas in PNe.  
As a result, 
issues such as the extent of turbulent mixing, the inhibition of 
heat conduction by magnetic fields, the ionization equilibrium of 
the hot material, or the origin of the X-ray-emitting material 
cannot be properly addressed.

To date, high-dispersion X-ray spectra have been reported 
only for BD+30$^\circ$3639, the X-ray brightest PN.  
Its detailed \emph{Chandra} Low Energy Transmission Gratings (LETG) 
high-dispersion X-ray spectrum \citep{Yu2009} has revealed many 
physical details and clarified conflicting descriptions of the 
chemical composition of its hot gas derived from CCD low-dispersion 
spectroscopic-imaging observations \citep{Maness_etal2003,Murashima2006}.  
The X-ray-emitting plasma has a temperature in the range
1.7$\times$10$^6$--2.9$\times$10$^6$ K, and the C and Ne 
abundances are unequivocally shown to be enhanced 
\citep{Yu2009}.  
This implies that the hot gas originated in nucleosynthesis processes that 
occurred deep in the AGB star, in the intershell where He-burning produced 
C.
Since the C- and Ne-rich gas has not mixed with nebular material, it
suggests that heat conduction and mixing between hot and cool gas do
not dominate the physical conditions of the X-ray-emitting plasma in
BD+30$^\circ$3639.

\begin{table*}[!t]
%\tablewidth{0pt}
%\tiny
\label{tab:obs}
\centering
\caption{\textit{XMM-Newton} observations details}
\begin{tabular}{cc|rr|rr|rr|rr}
\hline
\hline
Observation ID. & Rev. & \multicolumn{2}{c}{Raw exposure times} & \multicolumn{6}{c}{Net exposure times} \\ 
\cline{3-10}    &      & & & \multicolumn{2}{c}{High bck.} & \multicolumn{2}{c}{Moderate bck.} & \multicolumn{2}{c}{Low bck.} \\
                &      & RGS1  & RGS2  & RGS1  & RGS2  & RGS1  & RGS2  & RGS1  & RGS2  \\
                &      & [ks]~ & [ks]~ & [ks]~ & [ks]~ & [ks]~ & [ks]~ & [ks]~ & [ks]~ \\
\hline
0300570101      & 1143 &  20.1 &  20.1 & 16.75 & 11.29 &  5.65 &  0.59 &  2.55 & \dots \\ 
0300570201      & 1144 &  22.5 &  23.3 &  2.86 &  1.09 &  0.69 &  0.19 &  0.06 &  0.19 \\
0300570301      & 1145 &  22.3 &  22.2 &  4.49 &  8.75 &  0.06 &  0.73 &  0.06 &  0.73 \\
0300570401      & 1146 &  18.9 &  18.1 & 12.29 &  6.01 &  1.35 &  2.84 &  0.45 &  0.55 \\
0300570601      & 1141 &  10.9 &  11.1 &  5.43 &  6.90 &  4.62 &  3.73 &  3.13 &  2.14 \\
% 0300570701      & 1141 & \dots & \dots & \dots & \dots & \dots & \dots & \dots & \dots \\
0300570801      & 1142 &  14.3 &  14.1 & 11.21 & 10.50 &  5.32 &  3.83 &  3.17 &  1.98 \\
% 0300570901      & 1142 & \dots & \dots & \dots & \dots & \dots & \dots & \dots & \dots \\ 
% 0300571001      & 1143 & \dots & \dots & \dots & \dots & \dots & \dots & \dots & \dots \\ 
% 0300571101      & 1144 & \dots & \dots & \dots & \dots & \dots & \dots & \dots & \dots \\ 
% 0300571201      & 1145 & \dots & \dots & \dots & \dots & \dots & \dots & \dots & \dots \\ 
% 0300571301      & 1146 & \dots & \dots & \dots & \dots & \dots & \dots & \dots & \dots \\ 
0300571501      & 1161 &  21.6 &  21.6 & 17.85 & 20.19 & 12.58 &  8.84 &  4.94 &  2.11 \\
0300571701      & 1149 &  10.4 &  10.4 &  0.31 &  0.30 &  0.31 &  0.30 &  0.21 &  0.10 \\
% 0300571801      & 1149 & \dots & \dots & \dots & \dots & \dots & \dots & \dots & \dots \\ 
0300571901      & 1151 &  19.9 &  19.9 & 20.02 & 19.30 &  3.49 &  6.45 &  0.15 &  0.74 \\
% 0300572001      & 1151 & \dots & \dots & \dots & \dots & \dots & \dots & \dots & \dots \\ 
0300572101      & 1160 &  12.9 &  12.8 &  1.78 & 4.60  & \dots & \dots & \dots & \dots \\ 
% 0300572201      & 1160 & \dots & \dots & \dots & \dots & \dots & \dots & \dots & \dots \\ 
% 0300572301      & 1161 & \dots & \dots & \dots & \dots & \dots & \dots & \dots & \dots \\ 
0300572401      & 1162 &  13.4 &  13.4 &  0.06 &  0.06 & \dots & \dots & \dots & \dots \\
% 0300572501      & 1162 & \dots & \dots & \dots & \dots & \dots & \dots & \dots & \dots \\ 
0300572601      & 1163 &  14.5 &   7.9 &  3.31 &  2.22 & \dots & \dots & \dots & \dots \\
% 0300572701      & 1163 & \dots & \dots & \dots & \dots & \dots & \dots & \dots & \dots \\ 
0300572801      & 1164 &  19.4 &  19.3 & 11.64 & 17.86 &  6.35 &  3.17 &  1.03 &  0.74 \\
% 0300572901      & 1164 & \dots & \dots & \dots & \dots & \dots & \dots & \dots & \dots \\ 
\hline
TOTAL           &      & 221.1 & 214.2 &108.00 &109.07&  40.42 & 30.67 & 15.75 &  9.28 \\ 
\hline
\hline
\end{tabular}
\end{table*}

High-dispersion spectroscopy of the diffuse X-ray emission from PNe
would certainly represent a great advance in the derivation of 
physical conditions and chemical abundances of their hot gas, but it
requires high X-ray fluxes.  
The second X-ray brightest PN 
% There is presently only one additional PN that is bright enough 
for high-dispersion X-ray spectroscopic observations is NGC\,6543, 
a.k.a., the Cat's Eye Nebula. 
\emph{Chandra}'s superb angular resolution allowed \citet{Chu2001} to 
resolve the X-ray emission from NGC\,6543 into
a faint point-like emission source associated with the central star 
(CSPN) and a diffuse component within the central cavity.
% With the superb angular resolution of the \emph{Chandra} X-ray satellite,
% \citet{Chu2001} reported that the X-ray emission can be resolved into
% a faint point-like emission associated with the central star (CSPN)
% and a diffuse component within the central cavity.  
In this paper we
present high-dispersion X-ray spectra of NGC\,6543 obtained with the
Reflection Grating Spectrometers (RGS) on board \emph{XMM-Newton}.  The
observations and data reduction are presented in \S2. Special care
has been taken to minimize the effects of extended periods of high
background that compromised the data quality.  
The data have been used to gain insight into the physical conditions and 
chemical abundances of the hot gas inside this PN, as described in \S3.
The main results and conclusions are discussed in \S4.

\section{\textit{XMM-Newton} Observations}

The \emph{XMM-Newton} observations of NGC\,6543 consisted of 26 
individual pointings obtained in 2006 March and April (PI: M.A.\ 
Guerrero). 
The European Photon Imaging Cameras (EPIC) was used for all of 
them, but the RGS were used for only 14.  
In the following, we will focus our analysis in the RGS data.  
Details of the different observations, arranged by their observation 
ID, are listed in Table~1.  
% , where the observations not using the RGS are kept for completitude.  

The observations were reprocessed using the \emph{XMM-Newton} Science 
Analysis Software (SAS) 13.5.0 with the \emph{XMM-Newton} calibration 
files available in the Current Calibration File as of 2014 July 25.  
% EPIC and RGS event files were created using the SAS tools \textit{emproc}, 
% \textit{epproc}, and \textit{rgsproc}, respectively.  
RGS event files were created using the SAS tool \emph{rgsproc}.  In
order to identify and excise periods of high background from the
observations, light curves in the 10--12~keV energy range for the CCD
\#9 of each RGS observation were created. 
These light curves showed that most of the observations suffered from high 
background levels.  
This is the result of the visibility window of NGC\,6543 being close to the 
moment when the satellite approaches to its orbit perigee at the end of the 
revolution and to the observation dates in March and April.  
That orbital segment and seasonal period are known to exhibit the highest 
background levels \citep{XMM-Repo}.  
The situation was further aggravated by the high solar activity in 2006.  
Indeed, the high levels of background emission detected by the radiation 
monitor raised the warning flag active and commanded some of the five
\emph{XMM-Newton} instruments to the safe or standby mode for some of the 
26 observations initially planned.

\begin{figure}[!t]
\begin{center}
\includegraphics[bb=18 180 585 605,width=1.0\linewidth]{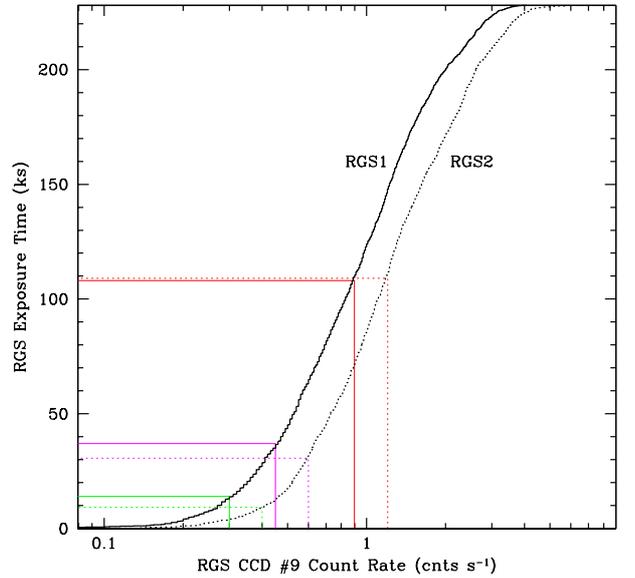}
\end{center}
\caption{
Cumulative frequency of the RGS1 (solid line) and RGS2 (dotted line) 
CCD \#9 count rate in the 10--12 keV energy range.  
The horizontal red, purple, and green lines mark the levels of 
CCD \#9 count rates adopted as high, moderate, and low background 
levels, respectively, for spectra extraction. 
}
\label{rgs.lc}
\end{figure}

The cumulative frequency of the background count rate of the RGS1 and 
RGS2 CCD \#9 is shown in Figure~\ref{rgs.lc}.  
We note that the RGS2 CCD \#9 count
rate is consistently a factor $\approx$4/3 higher than that of the RGS1 CCD
\#9.  In order to assess the effects of background on the quality of
the observations, we have extracted spectra for three levels of RGS1
and RGS2 CCD \#9 count rates: high ($\leq$0.9 cnts~s$^{-1}$ for RGS1
and $\leq$1.2 cnts~s$^{-1}$ for RGS2), moderate ($\leq$0.45
cnts~s$^{-1}$ for RGS1 and $\leq$0.6 cnts~s$^{-1}$ for RGS2), and low
($\leq$0.3 cnts~s$^{-1}$ for RGS1 and $\leq$0.4 cnts~s$^{-1}$ for
RGS2).  The net useful exposure times for individual spectra after
processing are listed in Table~1, and the final combined spectra, with
total exposure times of 217, 70, and 25 ks for the high-, moderate,
and low-background spectra, respectively, are shown in
Figure~\ref{rgs.spec}.

\section{Results and discussion}

\begin{figure}
\begin{center}
\includegraphics[bb=57 200 585 705,width=1.0\linewidth]{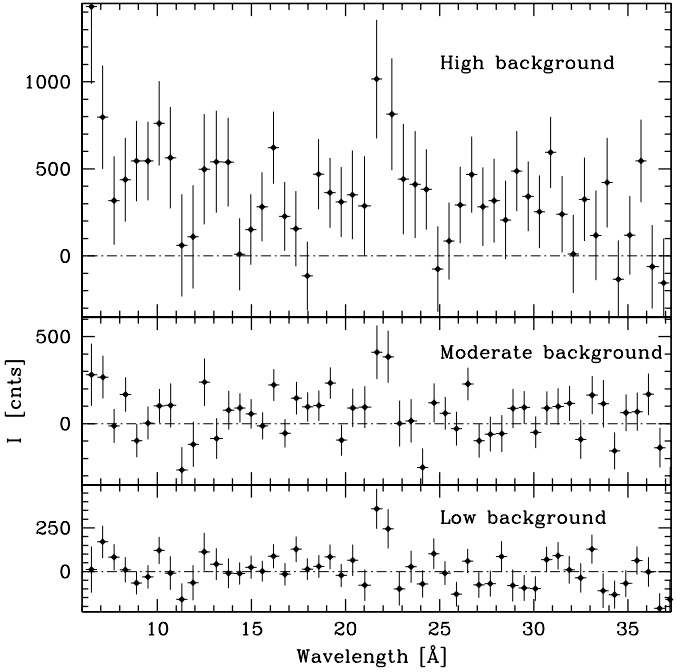}
\end{center}
\caption{
Combined background-subtracted RGS spectra of NGC\,6543 extracted for high 
(\emph{top}), moderate (\emph{center}), and low (\emph{bottom}) RGS1 and 
RGS2 CCD \#9 background count rates.  
The bin size is 0.30 \AA.  
}
\label{rgs.spec}
\end{figure}

An inspection of the spectra in Figure~\ref{rgs.spec} identifies consistently 
the presence of an emission feature at $\sim$22~\AA, which seems to correspond 
to the O~{\sc vii} triplet at 21.6, 21.8, and 22.1 \AA.  
The spectra show continuum emission whose level declines as the background 
level is reduced.  
For the last spectrum, with the most stringent background reduction, 
there is little or negligible continuum emission.  
The emission from the O~{\sc vii} triplet is detected at a 5-$\sigma$ 
level.

\begin{figure}[!t]
\begin{center}
\includegraphics[bb=45 230 585 705,width=1.0\linewidth]{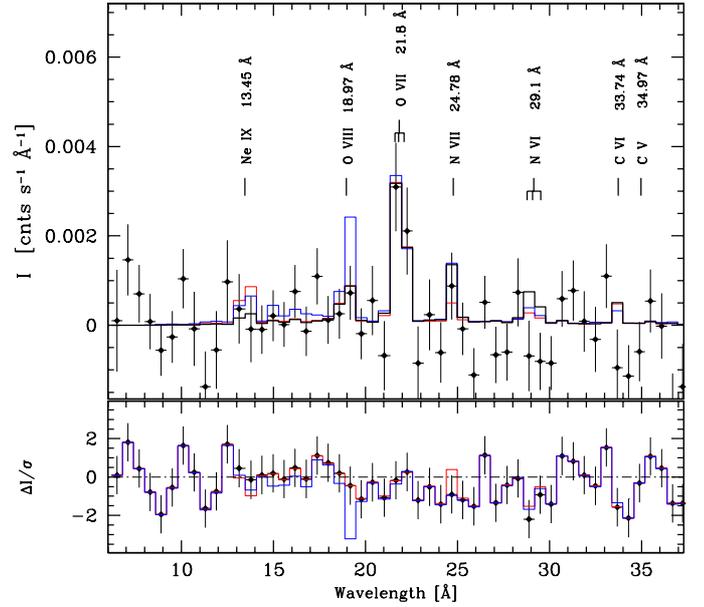}
\end{center}
\caption{
(\emph{top-panel}) 
Combined low-background (bottom panel of Fig.~2) background-subtracted 
RGS spectrum of NGC\,6543 overplotted by histograms of optically thin 
plasma emission models: 
black -- model with nebular abundances and a temperature of 0.147 keV, 
blue -- model with nebular abundances and a temperature of 0.19 keV, and 
red -- model with enhanced Ne and depleted N abundances and temperature 
0.147 keV.  
The intensity of each model has been normalized to that of the O~{\sc vii} 
triplet at 22 \AA. 
(\emph{bottom-panel}) 
Residuals of the models.  
The colors correspond to the same models as in the top panel.  
}
\label{rgs.mod}
\end{figure}

This spectrum is shown in more detail in Figure~\ref{rgs.mod}, where the 
wavelengths of emission lines that can be expected have been labeled.  
Besides the He-like lines of O~{\sc vii}, there is no significant evidence 
of the O~{\sc viii} 19.0 \AA, N~{\sc vi} 29 \AA, C~{\sc vi} 33.7 \AA, and 
C~{\sc v} 35.0 \AA\ lines, and only the N~{\sc vii} 24.8 \AA\ emission line 
might have been detected at a low, questionable confidence level.  
When compared with the \emph{Chandra} high resolution LETG spectrum of 
BD+30$^\circ$3639 \citep{Yu2009}, NGC\,6543 shows also the O~{\sc vii} 
triplet, but the bright Ne~{\sc ix} 13.45 \AA\ line and the relatively 
intense H-like lines of O~{\sc viii} and C~{\sc vi} in the former are 
completely absent in NGC\,6543.  
These spectral variations imply notable differences in the physical 
conditions and/or chemical abundances of the X-ray-emitting plasma 
between these two PNe.

Obviously, the quality of the RGS spectrum of NGC\,6543 is not suitable 
for deriving the physical conditions and chemical abundances of its hot gas 
by means of spectral fits.  
Instead, these will be investigated by comparing qualitatively the RGS 
spectrum with the spectra computed from different optically-thin plasma 
emission models.  
For this purpose, the XSPEC V12.8.2 \emph{apec} code has been used.  
% First, the high helium abundances (He/He$_\odot$=60) derived for the 
% stellar wind of its CSPN \citep{deKoter96} can be completely discarded, 
% as it implies higher levels of continuum emission than line emission.  
The chemical abundances of the emitting plasma have been assumed to be those 
of the nebula reported by \citet{BernardSalas2003}, i.e., C/C$_\odot$=0.9, 
N/N$_\odot$=3.4, O/O$_\odot$=1.1, and Ne/Ne$_\odot$=2.2 \citep{Asplund09}.  
These nebular abundances are otherwise similar to those of the stellar 
wind of the CSPN derived by \citet{Georgiev2008}, whereas the high 
helium abundances (He/He$_\odot$=60) of the stellar wind suggested by 
\citet{deKoter96} can be completely discarded, as it would imply higher 
levels of continuum emission than line emission.  
Then, the hydrogen column density was set to $N_\mathrm{H}$=4.1$\times$10$^{20}$ 
cm$^{-2}$, in accordance with its color excess $E_{B-V}$=0.07 
\citep{BernardSalas2003}.  
The plasma temperature implied by \citet{Chu2001}, $kT$=0.147 keV 
(=1.7$\times$10$^{6}$~K), results in the synthetic spectrum shown in 
Figure~\ref{rgs.mod} (black histogram).

% ignore 2.0-**
% setpl reb 5 60  
% 3.2E-4 == 2.0e-13 erg/cm/cm/s between 0.3 and 2.0 keV

The intensity predicted for different emission lines is generally
consistent with the observed spectrum.  The nondetection of the
O~{\sc viii} 19.0 \AA\ implies an upper limit of 0.19~keV for the
plasma temperature, as constrained by the 3-$\sigma$ level of the line
intensity (blue histogram in Fig.~\ref{rgs.mod}).
% 3.0e-4, 0.19 keV
The C and Ne abundances are also consistent with the observed spectrum, 
but it should be noted that an increase in the Ne abundances by a 
factor of 4 and a depletion of the N abundances by a factor of 3 would 
still produce acceptable results (red histogram in Fig.~\ref{rgs.mod}). 
The conspicuous absence of the N~{\sc vi} 29.1 \AA\ triplet may hint 
at a solar abundance for N, as the alternative way to diminish the line 
intensity by raising the plasma temperature is constrained by the upper 
limit in temperature imposed by the O~{\sc vii} to O~{\sc viii} line 
ratios.  
Moreover, the absence of emission from this triplet sets a lower limit 
for the plasma temperature not much lower than the temperature of 0.147 
keV used in our main model.

\begin{figure}[!t]
\begin{center}
\includegraphics[bb=57 240 585 705,width=1.0\linewidth]{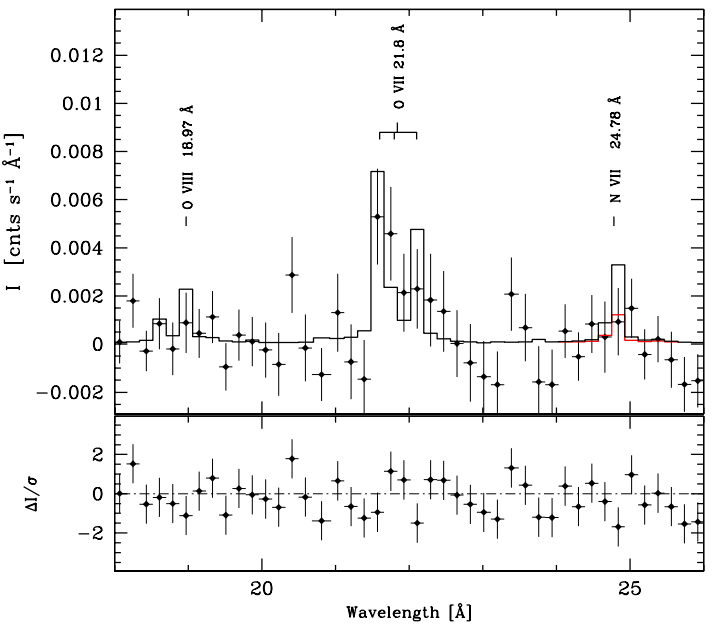}
\end{center}
\caption{
Same as previous figure, but for a bin size of 0.09 \AA.  
The O~{\sc vii} triplet is resolved in this spectrum.  
The black histogram corresponds to an optically thin plasma emission 
model with nebular abundances and a temperature of 0.147 keV.  
The red histogram, as in the previous figure, considers a solar 
N/O ratio. 
}
\label{rgs_bin18}
\end{figure}

A higher resolution close-up of the spectrum is shown in
Figure~\ref{rgs_bin18}.  In this spectrum, the components of the
O~{\sc vii} triplet are resolved.  
Within the uncertainties, the relative intensities of the O~{\sc vii} 
21.60 \AA\ resonance (\emph{r}), 21.80 \AA\ intercombination (\emph{i}), 
and 22.10 \AA\ forbidden (\emph{f}) lines are consistent with those of a 
coronal plasma, as expected for the low densities of the plasma inside 
the hot bubbles of PNe. 
The nondetection of the H-like lines of N~{\sc vii} at 24.78 \AA\ and 
O~{\sc viii} at 18.97 \AA\ is noticeable.  
% It is significant the conspicuous absence in this spectrum of emission 
% corresponding to the H-like lines of N~{\sc vii} at 24.78 \AA\ and 
% O~{\sc viii} at 18.97 \AA.  
The lack of O~{\sc viii} emission
sets a reliable upper limit for the plasma temperature, but this is
problematic for the Ly-$\alpha$ line of N~{\sc vii}.  A plasma
temperature lower than the adopted value of 0.147 keV would imply a
brighter N~{\sc vi} 29.1 \AA\ triplet, which is completely absent.
This seems to indicate that the actual abundance of N in the
X-ray-emitting plasma is lower than those used in our model,
N/N$_\odot$=3.4. For comparison, a model with solar N/O ratio is
also shown (red histogram in Fig.~\ref{rgs_bin18}).

The plasma temperature constrained by the \emph{XMM-Newton} RGS spectrum 
is consistent with that reported in previous studies of the diffuse X-ray 
emission of NGC\,6543 based on \emph{Chandra} ACIS data 
\citep{Chu2001,Maness2003}.
The nondetection of the
H-like line of O~{\sc viii}, contrary to the case of
BD+30$^\circ$3639, implies a lower temperature for the X-ray-emitting
plasma of NGC\,6543. There is no evidence for an enhancement of the C
and Ne abundances similar to those derived for BD+30$^\circ$3639 using
\emph{Chandra} LETG \citep{Yu2009}; the C/O ratio of 15-45 found in
BD+30$^\circ$3639 is not seen in NGC\,6543, but Ne abundances several 
times solar can not be completely excluded.

The RGS spectrum of NGC\,6543 sheds some light on the disputed N/O
ratio of the X-ray-emitting plasma in this nebula
\citep{Chu2001,Maness2003}.  The weak or absent N~{\sc vii} and N~{\sc
  vi} emission lines implies low N abundances.  The N/O ratio is
apparently closer to the ratio of 1.65$\times$N$_\odot$/O$_\odot$
found for the stellar wind \citep{Georgiev2008} than to the ratio of
3.4$\times$N$_\odot$/O$_\odot$ found in the nebula
\citep{BernardSalas2003}.  We caution, however, that the low
signal-to-noise ratio of the RGS spectrum does not allow us to obtain
an accurate estimate of the N/O ratio.

\section{Conclusions}

We have presented \emph{XMM-Newton} RGS observations towards the Cat's
Eye Nebula, NGC\,6543.  The total RGS exposure time was 435.3~ks, but
the observations were severely affected by high-energy background
episodes that reduce significantly the quality expected for this data
set.  Different levels of background emission have been considered and
analyzed to make the best use of the data. Unfortunately, only the
most stringent levels of background emission allow a reliable analysis
of the spectrum of NGC\,6543. In the end, the net useful exposure time of the
spectrum used for analysis is 25 ks, i.e., only a meager $\sim$6\% of
the total exposure time.

This is the second PN reported with high resolution spectroscopic
observations apart from BD\,$+$30$^{\circ}$3639 \citep{Yu2009}.  The
final background-subtracted combined RGS spectrum only shows a clear
detection of the He-like triplet of O~{\sc vii} at $\approx$22 \AA,
whereas the N~{\sc vii} 24.78 \AA\ line is only tentatively detected.
The weakness or absence of the H-like O~{\sc viii} line points to
lower X-ray temperatures than in BD+30$^\circ$3639.  Indeed, we find
the X-ray temperature to be consistent with that derived from the
\emph{Chandra} ACIS low-resolution spectrum, $kT\approx0.147$ keV 
or $T\approx1.7\times10^6$ K
\citep{Chu2001}. Otherwise, the weakness or absence of Ne and C lines
indicates much lower Ne/O and C/O than those of BD+30$^\circ$3639
derived from its LETG spectrum. 
The weakness of the N lines suggests a low N/O ratio as well, although 
we note the data quality is not sufficient for a solid statement.

The laborious X-ray analysis of the \emph{XMM-Newton} RGS
observations of NGC\,6543 and the low quality of this data set 
% make evident that we have reached the technical limits 
unveil the difficulties for the X-ray study of PNe at high dispersion. 
Hopefully, \emph{Astro-H} and \emph{Athena} will be able to extend 
in the future \citep{Nandra2013,Takahashi2012} 
the deep \emph{Chandra} and \emph{XMM-Newton} grating observations 
of the X-ray brightest PNe.  
In particular, the Soft X-ray Spectrometer (SXS), a microcalorimeter on 
board \emph{Astro-H}, to be launched next year, may soon provide us 
with improved-quality high-dispersion X-ray spectra of PNe.  

% similar efective area as the combined \emph{CXMM-Newton} RGS, 
% but with lower background  will count with a low-noise, high-resolution 
% micro-calorimeter which can provide 

\acknowledgements 
J.A.T.\ and M.A.G.\ are partially funded by AYA 2005-01495 of the Spanish MEC 
(Ministerio de Educacion y Ciencia) and AYA 2011-29754-C03-02 of the 
Spanish MICINN (Ministerio de Ciencia e Innovaci\'on) co-funded with 
FEDER funds. 
J.A.T.\ acknowledges support by the CSIC JAE-Pre student grant 2011-00189. 
Y.-H.C.\ and R.A.G.\ acknowledge the support of the NASA \emph{XMM-Newton} 
Guest Observer grant NNG06GE67G for this research project.
We thank the referee, J.H.\ Kastner, for his helpful comments.

\end{document}